\documentstyle[12pt,epsf,aasms4]{article}
\begin{document}

\title{
Observations of TeV gamma ray flares from Markarian 501\\
with the Telescope Array Prototype
}

\author{
N.Hayashida, 
H.Hirasawa, 
F.Ishikawa, 
H.Lafoux, 
M.Nagano, 
D.Nishikawa, 
T.Ouchi, 
H.Ohoka, 
M.Ohnishi,
N.Sakaki, 
M.Sasaki, 
H.Shimodaira,
M.Teshima,
R.Torii,
T.Yamamoto, 
S.Yoshida, 
and T.Yuda
}

\affil{
Institute for Cosmic Ray Research, University of Tokyo, Tokyo 188, Japan.}

\author{
Y.Hayashi, 
N.Ito, 
S.Kawakami, 
Y.Kawasaki, 
T.Matsuyama, 
M.Sasano, 
and T.Takahashi
}

\affil{Department of Physics, Osaka City University, Osaka 558, Japan.}

\author{
N.Chamoto,
F.Kajino, 
M.Sakata, 
T.Sugiyama, 
M.Tsukiji,
and Y.Yamamoto
}
\affil{Department of Physics, Konan University, Kobe 658, Japan.}

\author{
N.Inoue,
E.Kusano, 
K.Mizutani, 
and A.Shiomi
}
\affil{Department of Physics, Saitama University, Urawa 338, Japan.}

\author{
K.Hibino, 
T.Kashiwagi, 
and J.Nishimura
}
\affil{Department of Engineering, Kanagawa University, Yokohama 221, Japan.}

\author{
E.C.Loh, 
P.Sokolsky, 
and S.F.Taylor
}
\affil{Department of Physics, University of Utah, USA.}

\author{
K.Honda, 
N.Kawasumi, 
and I.Tsushima
}
\affil{Faculty of Education, Yamanashi University, Kofu 400, Japan.}

\author{Y.Uchihori}
\affil{The 3rd Research Group, National Institute of Radiological Sciences, Chiba 263, Japan.}

\author{H.Kitamura}
\affil{Department of Physics, Kobe University, Kobe 657, Japan.}
\author{M.Chikawa}
\affil{Research Institute for Science and Technology, Kinki University, Osaka 577, Japan.}
\author{S.Kabe}
\affil{National Laboratory of High Energy Physics, Tsukuba 305 ,Japan.}
\author{Y.Mizumoto}
\affil{National Astronomical Observatory, Tokyo 181, Japan.}
\author{Y.Matsubara}
\affil{Solar-Terrestrial Environment Laboratory, Nagoya University, Nagoya 464-1, Japan.}
\author{H.Yoshii}
\affil{Faculty of General Education, Ehime University, Ehime 790, Japan.}
\author{N.Hotta}
\affil{Faculty of Education, Utsunomiya University, Utsunomiya 321, Japan.}
\author{To.Saito}
\affil{Tokyo Metropolitan Colleage of Aeronautical Engineering, Tokyo 116, Japan.}
\author{M.Nishizawa}
\affil{National Center for Science Information System, Tokyo 112, Japan.}
\author{K.Kuramochi}
\affil{Informational Communication, KOKUSAI Junior College, Tokyo 165, Japan.}
\author{K.Sakumoto}
\affil{Faculty of Science and Technology, Meisei University, Tokyo 191, Japan.}

\authoremail{mteshima@icrr.u-tokyo.ac.jp}

\begin{abstract}
We will report the observations of TeV gamma ray flares from Markarian 501
using Telescope Array Prototype. 
The observation were carried out continuously from the end of
March to the end of July in 1997. 
The energy spectrum, and the time variation of
the gamma ray intensities are shown. The intensity has been changed
by the order of magnitude in this period and the possible quasi periodic 
oscillation of 12.7days were discovered.
\end{abstract}


\section{Introduction}

Markarian 501 (Mkn501) is an extragalactic BL-Lac type object at z=0.034.
It is observed in radio, optical and X-ray bands, and can be
characterized in flat spectrum radio, 
highly optically polarized and optically violently variable.
BL-Lac type objects are considered to be a kind of AGN, 
which have a jet oriented to our line of the sight.
Recently, Mkn501 (\cite{Whipple-501,HEGRA-501}), 
Mkn421 (\cite{Whipple-421,HEGRA-421,TA-421}), 
and 1ES 2344+514 (\cite{Whipple-1ES}) have been identified as TeV
gamma ray sources, and they are all BL-Lac objects.
Gamma rays from Mkn501 around 100MeV to 10GeV
was not observed by the CGRO EGRET detector(\cite{EGRET}), 
however, emission in the TeV energy region were discovered 
by the Whipple telescope (\cite{Whipple-501}) and
confirmed by the HEGRA telescopes (\cite{HEGRA-501}).
Detailed study of gamma rays from AGN will give us
information about the environment surrounding the huge black hole
located at the center of the AGN, and the high energy phenomena
and particle acceleration in the jet.
The time scale of the intensity variations of the TeV gamma rays 
may explain the particle acceleration site is close to 
the black holes.
TeV gamma rays from such extragalactic object interacts with
the infrared background photons, and a cutoff of the energy spectrum
is expected around 7-15TeV(\cite{stecker}). 
There is an ambiguity in the prediction
of the cutoff energy due to the uncertainty of the number density of 
infrared photons.
In the discovery stage, the intensity in the TeV energy range was very small,
corresponding to 8\% of the Crab nebula flux, however, from Mar 1997 
it increased dramatically and variated from 0.3  
to 4 Crab flux. It is found that the variation of the intensities 
are larger in the TeV range than X-ray and other ranges(\cite{Whipple-multi}).
These flares are observed by Whipple , 
HEGRA, CAT, TACTIC and the Telescope Array Prototype (TAP) (\cite{Whipple-501,HEGRA-501,CAT-501,TACTIC-501,TA-501}). 
The flares were continuously observed until the end of July 1997.
Here we will report the details of the Mkn501 observations by TAP.

\section{Experiment}

The Telescope Array Prototype detector \cite{TA-array}, 
a seven telescope array is 
under construction at Utah, Dugway. Its geographical position is
$40.33^\circ$ N, $113.02^\circ$ W at an altitude of 1600m above sea level.
The prototype detector works in dual modes: 
the Cherenkov and the air fluorescence mode. 
Construction started 
in the summer of 1996, and currently three telescopes with the respective
separation of 120m are in operation.
The seven telescopes will be arranged in a hexagonal grid 
with a separation of 70m to maximize the detection efficiency of 
TeV gamma rays. 
Each telescope is the alt-azimuth mount with a 6m$^2$ 
main dish. The main dish consists of 19 hexagonal shape segmented
mirrors coated with anodized aluminum. The reflectivity of mirrors 
are about 90\% at wave length of 400nm. At the focal plane, 
a high resolution imaging camera of 256ch photomultipliers 
is installed to measure detailed images of Cherenkov light from
gamma rays and cosmic rays. The Cherenkov light images are 
used to distinguish the gamma rays from the huge number 
of background cosmic rays.
The typical cosmic-ray rate is about 1000/min and the gamma-ray rate
from the Crab nebula is about 0.5/min (\cite{TA-501}) with three telescopes. 
Therefore rejection 
of cosmic ray background events using the shape parameter of the Cherenkov 
light is essential to obtain a reasonable S/N ratio in this experiment, and we
use the techniques originally developed by the Whipple group 
(\cite{Whipple-image}).
The absolute pointing accuracy of the telescopes is typically 1 arcmin
which is frequently calibrated by imaging bright stars.

The signals from the 256ch camera are amplified just behind the camera
in order to minimize electric noise and to obtain better timing
resolution, and are then fed to ADC and TDC modules mounted at the telescope
base to measure the amplitude and the timing after passing through 
10m of twisted pair cables.
The triggering requirement to record the event is four folds out of 256 
tubes. The threshold of the discriminators are set at 5 photoelectron
level. The single counting rates in each tube are set to 3-5kHz 
in each channel.
The threshold energy for detectable gamma rays is 600 GeV for 
vertical showers.
The Mkn501 was observed with the raster scan tracking mode.
In this mode, the telescope tracking center scans
the sqaure region of $\pm 0.5 deg$ in right ascension and declination
coordinate centered on the target.
There are several advantages in this method compared with the
conventional on-source / off-source tracking mode.
The on-source and off-source sky region can be observed simultaneously.
The systematics of the imaging devices can be reduced significantly.
By observing the bright star images, the calibration of the telescope 
absolute direction can be done with the accuracy of $0.03^\circ$.
Our telescopes will be described in more detail elsewhere.

This observation of Mkn501 was carried out from the end of March to
the end of July, 1997. We observed a total number of 47 nights for a 
total observation
time of 105.4 hrs. We have observed 3,400,000 events in the F.O.V. of 
$\pm 2^\circ$ around Mkn501, which are mainly cosmic-ray protons and 
Helium nuclei. 
In the analysis, we have limited the zenith angle to the range of
$5^\circ$ to $25^\circ$ in order to reduce the systematic errors in the
energy and the aperture estimates. This leaves 2,160,000 events and 64.0 hrs
of live time, with an average event rate of 9.4Hz with three telescopes.
Among these events, we selected gamma ray 
candidates using the shape parameters of Cherenkov images and 
their directionality. 
We can determine the arrival direction of gamma rays and the cosmic rays
with an accuracy of $ 0.1^\circ \times  0.3^\circ$ 
(the elliptical errors) in each event. Therefore, 97\% of the background
events around the target, say within $\pm 1^\circ$, can be rejected. 
The cosmic rays show larger images than gamma rays, and they are 
rejected with 95\% efficiency through the image selection.
Therefore, typically 99.95 \% of the background events around the target
are rejected with the image and the arrival direction 
using the present analysis method.
35 \% of the gamma rays remain through this process.
Therefore, even if the original $S/N$ is bad as $0.001$,
we can obtain $S/N$ of order $\sim 1$.

\section{Analysis}

At first, the clean Cherenkov images are obtained by removing the
background photons with the timing information.  
The signal timings of each photomultipliers are recorded with
an accuracy of 1nsec.
The Cherenkov light signals from gamma rays and cosmic rays 
are concentrated within a 10 nsec interval, however
noise including the photons from star light and the air glow 
are randomly distributed in time. Therefore we require a timing 
alignment with a software gate set at 40 nsec. 
Then we require the clustering of hit tubes, geometry.
After these selections, the chance coincidence events due to the random
coincidence are completely rejected and the pure Cherenkov events 
induced by gamma rays and cosmic rays remain.

Then, the image parameters are calculated: signal size SIZE, 
the centroide position of
images (x,y), WIDTH, LENGTH and CONC.
The events with images located near the camera boundary
are cut by requiring the condition $R= \sqrt{x^2+y^2} \le 1.8^\circ$,
because the images near the camera boundary are distorted 
due to the boundary.
The gamma rays show the compact images compared with the
hadronic showers, so we then selected the events which
has a smaller WIDTH and LENGTH region similar to what is predicted by a 
Monte Carlo simulation.  We select events with the following conditions,
$WIDTH \le (W_{30}+0.020 \times ln(SIZE/400))$, 
and $ LENGTH \le L_{50} + 0.023 \times ln(SIZE/400)$,
where $W_{30}$ and $L_{50}$ correspond to 30\% and 50\% of the
points obtained by integrating the width distribution 
from off-source events (Hadron events).
The typical values of $W_{30}$ and $L_{50}$ are $0.15^\circ$ and $0.35^\circ$,
respectively, and they naturally contain the
zenith angle dependencies and the weather conditions,
and we could obtain the constant fraction of data as
gamma ray candidates. The SIZE of 400 corresponds to the the average 
value of SIZE. 
The second terms (0.020 and 0.023) represent
the SIZE dependence of the images as predicted by
the Monte Carlo simulation.
The parameter $CONC \ge C_{50}$ corresponds to the
light concentration, as gamma ray showers have a higher light concentration
than hadronic showers.
After these selections, 8-10\% of events remain.
Finally, we select for gamma ray showers using the 
directional information obtained by the asymmetry of shower
images. We select for events which develop from neighbouring of the target source to outer direction.
Finally, 97\% of the hadronic showers are rejected and
30-40\% of gamma rays are selected. 
After these cuts we obtained the excess of $35 \sigma$
from the direction of Mkn501. The monthly excesses in the alpha distribution 
are shown in Figure 1.

\subsection{Energy Spectrum}

The absolute gain calibration is carried out by the measurement of
a single photoelectron. We found
that a single photoelectron corresponds to 4 ADC counts.
The relative gain of each channel of 256ch is adjusted to within 5\%
accuracy using the LED pulsar.
We confirmed that the $SIZE$ distribution observed 
is consistent with the simulated showers assuming the cosmic ray energy
spectra in each composition. The uncertainty of the
absolute gain or the $SIZE$ and the energy relation is
estimated as to be 20\%.
With a Monte Carlo simulation, we can determine the effective area
for gamma rays and hadrons, $S_{g}(SIZE)$ and 
$S_{h}(SIZE)$. We also obtain experimentally the number
of the excess events $N_{ex}(SIZE)$, and
number of background events $N_{b.g.}(SIZE)$. 
In order to minimize the systematic errors, we took
the ratio $R(SIZE) = (N_{ex}(SIZE)/N_{b.g.}(SIZE))/
(S_{g}(SIZE)/S_{h}(SIZE))$.
The denominator $S_{g}/S_{h}$ has a weak dependence
on SIZE with our imaging selection of approximately $SIZE^{0.4-0.5}$.
The energy and the $SIZE$ relations for gamma rays and hadrons 
are determined by our Monte Carlo simulation to be
$E_g = SIZE/300$ TeV and $E_h = SIZE/100$ TeV, respectively.
Then we can obtain the gamma ray energy spectrum from
$ dF/dE \propto R(SIZE) \times (dF/dE_h)$.
The derived energy spectrum is shown in the Figure 2.
$dF/dE = (4.0 \pm 0.2) \times (E/1TeV)^{-2.5 \pm 0.1}$.
The differential energy spectrum can be well fit with 
the power law spectrum of index -2.5 up to 5 TeV.
The spectrum becomes steeper above 5TeV which may suggest
a cutoff of the energy spectrum. However, it is possible that
statistical fluctuations may make this effect. We need more statistics
to obtain conclusive results.
The saturation of the photomultipliers, the amplifiers, and
the ADC is considered to occur at higher energy, above 
30 TeV.  We could minimize this saturation effect by the above ratio method.

\subsection{Time Variation}

The observation of the Mkn501 was carried out for 47 nights
from the end of March to the end of July.
The gamma ray event rate is plotted as a function of MJD 
in Figure 3. For reference, the gamma ray rate from the Crab Nebula
measured by our detectors is shown by a horizontal line.
The event rate was highly variable day by day, and
the maximum event rate was about 4 Crab and the minimum
rate was $0.3\pm0.3$ Crab.
The time scale of the intensity change was about a few days.
We searched for a short time variation of the intensity
but we could not find clear evidence of any short time variation 
in our data set. 

We can see high states and low states clearly in our data set.
This feature (the time scale and the intensity change) in April and in July 
appear to be similar, showing ``U'' shapes, and the interval of
the two high states are 14 days and 12 days. May and June
data each show only one high state ``$\Lambda$'' shapes.
In order to examine the periodicity in the data set,
the data shifted by 25.5 days are superposed in the bottom pannel of
Figure 3. We can see good coincidence of the high states.
This test suggests a $25.5 \pm 2$ day periodicity.

The periodicity of the gamma ray intensities was examined
by the Rayleigh test. We calculated the powers (amplitude when fitting
sinusoidal function) in each test period from 5 days to 45 days.
In order to evaluate the significance in each test period, we have
generated $10^7$ data set artificially by shuffling the relation of
($I_g$, MJD), where $I_g$ is 47 flux points in Figure 3. 
(These data set also has the same gaps in the observation time.)
They are analyzed in the same way as the real data set.
By comparing the amplitude of the real data set with 
the amplitude distribution of the generated $10^7$ data sets, 
we can estimate the probability of the appearance of observed amplitude
in each test period.
With this method we could remove the spurious effect (the gaps in the
observations making the false periodicities), and 
succeed to obtain the chance probability.
Figure 4 shows the obtained chance probability 
as a function of the test period.
We can see prominent peak at 12.7 days with the chance probability of less
than $10^{-5}$. This period corresponds to a half of the period obtained in
the simple test mentioned above.

\section{Discussion}

We obtained the differential energy spectrum of the gamma rays from
Mkn501. It shows a possible cutoff feature above 5-7 TeV. This cutoff
is predicted to be caused by the interaction with the infrared photons or
the the limit of the electron acceleration in the jet.
The cutoff energy of the gamma rays due to the
infrared photon interaction from the distance of the Mkn501, z=0.034 
is estimated 7 TeV to 15 TeV by \cite{stecker}. 

In the time variation of the gamma ray intensities, we found
a periodicity, which may be a quasi periodic oscillation.
The observed periodicity corresponds to 12.7 days or 25.5 days.
The relationship of this periodicity with high energy phenomena around
the massive black holes could be influenced by factors including
the precession of the jet, and the rotation of the black hole(s).
R. Protheroe (\cite{Ray}) suggested the interaction of the shock wave and
the helical structure of the jet may cause this observed type of periodicity.

\acknowledgments

Authors would like to thanks to J.Como, A.Larsen, R.Smith, F.Misak, 
L.Sutton, D.Gardener, S.Davis, C.Davis, J.Parry and B.Larsen 
for their contribution to our experiment in USA.
This project is supported by the grant aid(\# 07247102 )
and aid(\# 08041096 ) by MONBUSHOU (the ministry of education and science)
and also partly supported by the Science Research Promotion Fund  
of Japan Private School Promotion Foundation.

\begin{figure}[h]
\begin{center}
\plotone{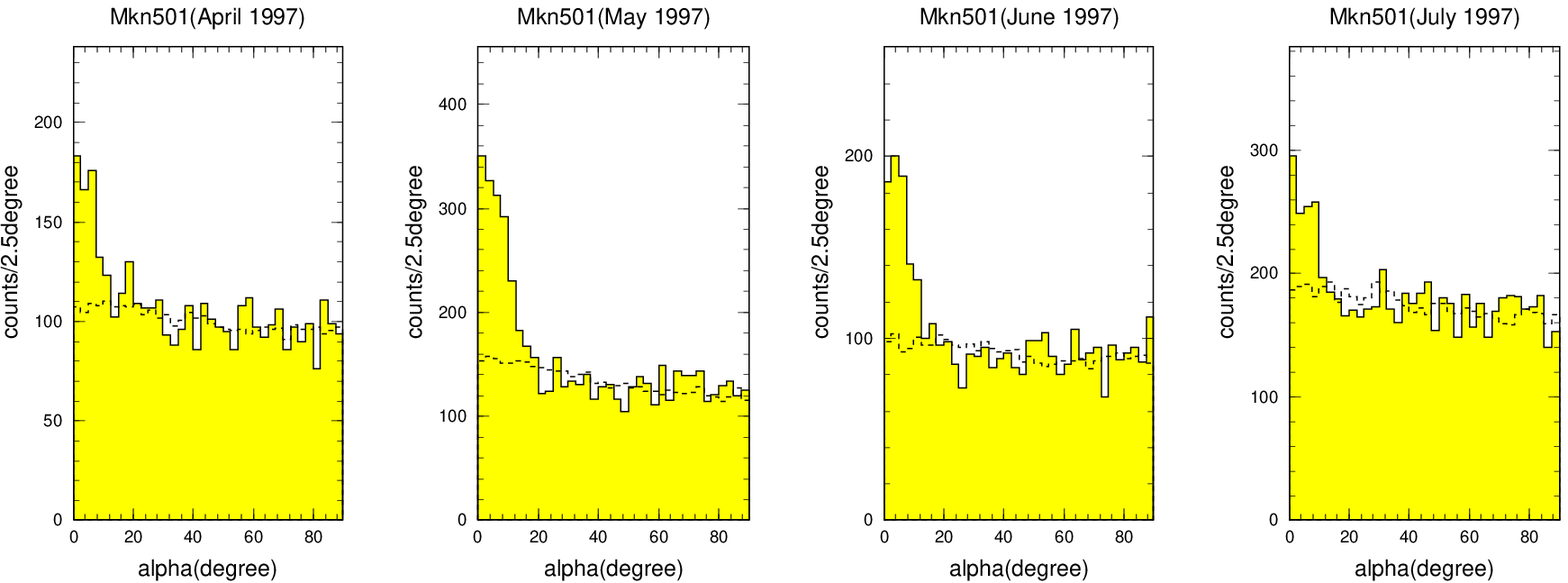} 
\caption{
	The monthly alpha distributions.
        Solid lines and broken lines show the alpha distribution for on-source,
        and for off-source, respectively.
        The excess in the small alpha region ($\le 15^\circ$) corresponds
        to the gamma rays from Mkn501.
        The total significance is more than $35 \sigma$.
	}
\end{center}
\end{figure}

\begin{figure}[h]
\begin{center}
\epsscale{0.7}
\plotone{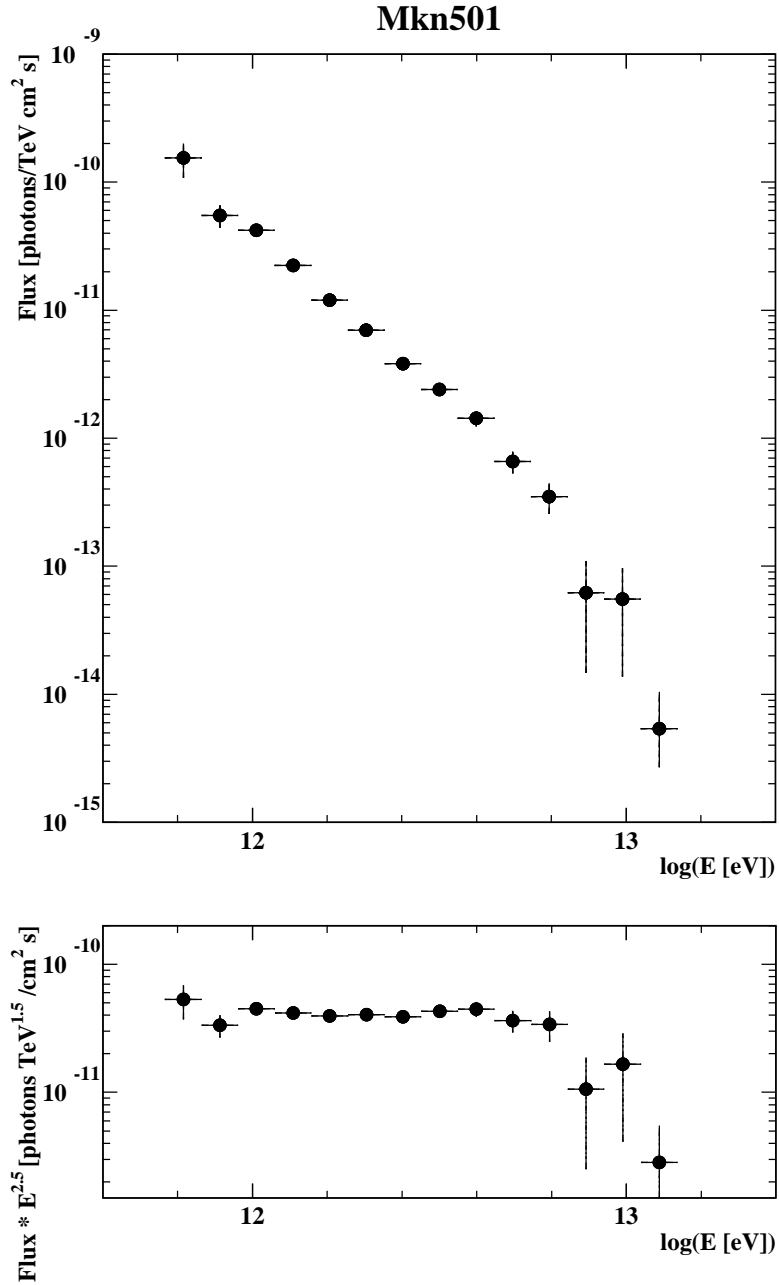} 
\caption{
	The top pannel shows the differential energy spectrum of 
        gamma rays from Mkn501. The bottom one is multiplied by $E^{2.5}$ to
        see the detail structure of the spectrum.
	}
\end{center}
\end{figure}

\begin{figure}[h]
\begin{center}
\plotone{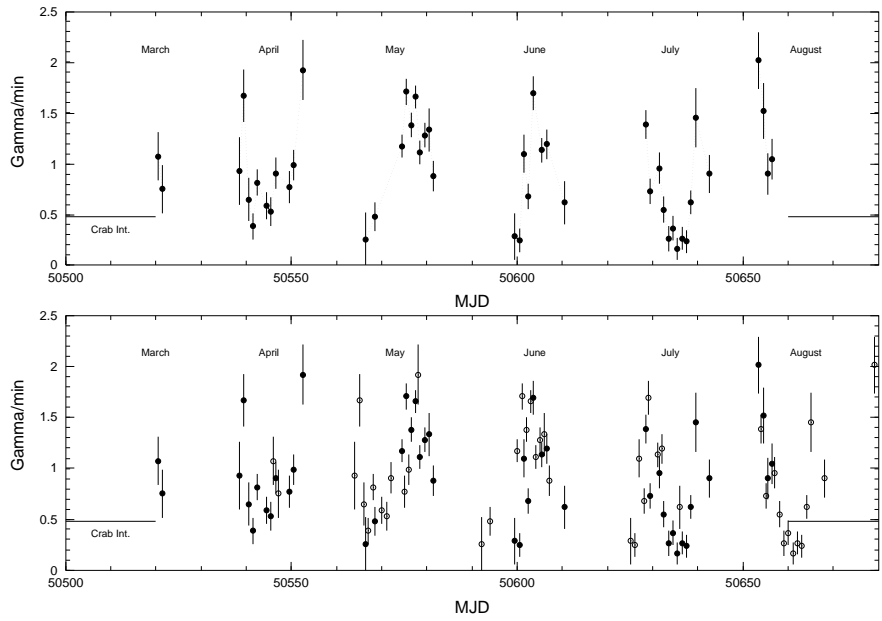} 
\caption{
	The top pannel shows the time variation of the intensity of gamma rays 
        from Mkn501.
        In the bottom panel, the data shifted by 25.5 days are superposed to
        the original data. We see good agreement between the original
        data and the artificially shifted one.
	}
\end{center}
\end{figure}

\begin{figure}[h]
\begin{center}
\plotone{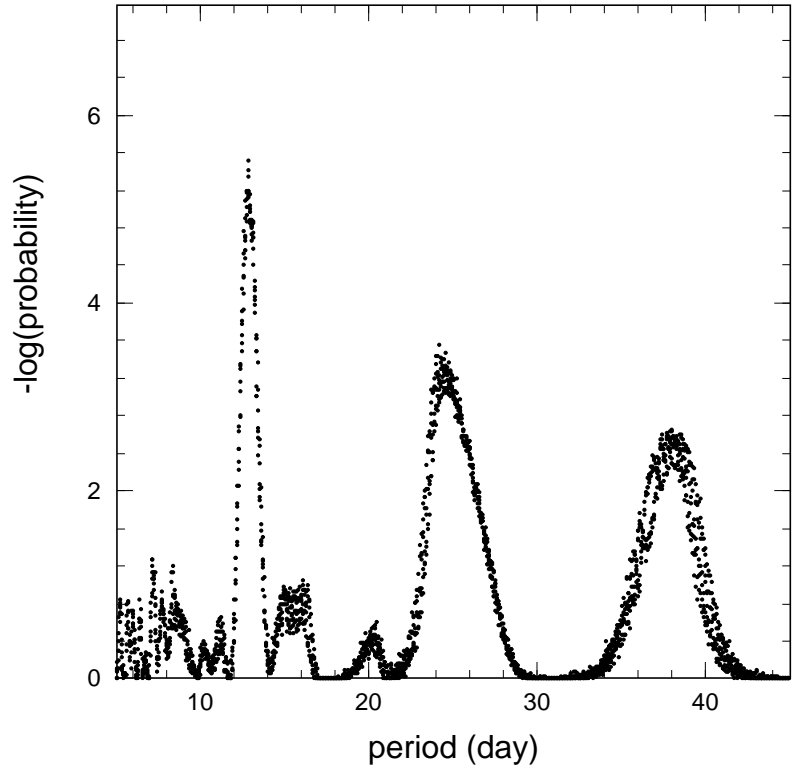} 
\caption{
	The periodicity test. The Rayleigh tests are carried out
        assuming the test period. The amplitude is calculated in each
        test period and the occurance of the amplitude is evaluated
        using $10^7$ false data sets.
	}
\end{center}
\end{figure}

\end{document}